\documentclass[a4paper]{article}
\usepackage{amsmath}
\usepackage{authblk}
\usepackage[usenames,dvipsnames,svgnames,table]{xcolor}
\usepackage{graphicx}
\usepackage{subcaption}
\usepackage{lineno}
\usepackage{mathrsfs}
\usepackage{longtable}
\usepackage{multirow}
\usepackage{hyperref}
\usepackage[a4paper,margin=1in]{geometry}
\usepackage[utf8]{inputenc}
\newcommand*\keywords[1]{%
  {\renewcommand{\baselinestretch}{0.95}\baselineskip 12pt
     \noindent{\bf\quad #1\par}}}
\begin{document}
\title{Signal-background discrimination with convolutional neural networks in the PandaX-III experiment using MC simulation}
\author[1]{Hao Qiao}
\author[2]{Chunyu Lu}
\author[2]{Xun Chen\footnote{Corresponding author: chenxun@sjtu.edu.cn}}
\author[2]{Ke Han}
\author[3,2]{Xiangdong Ji}
\author[1]{and Siguang Wang
\footnote{Corresponding author: siguang@pku.edu.cn}}

\affil[1]{School of Physics and State Key Laboratory of Nuclear Physics and Technology and Center for High Energy Physics, Peking University, Beijing 100871, China}
\affil[2]{INPAC and School of Physics and Astronomy, Shanghai Jiao Tong University, Shanghai Laboratory for Particle Physics and Cosmology, Shanghai 200240, China}
\affil[3]{T.D. Lee Institute, Shanghai 200240, China}
\maketitle

\abstract{The PandaX-III experiment will search for neutrinoless
  double beta decay of $^{136}$Xe with high pressure gaseous time
  projection chambers at the China Jin-Ping underground
  Laboratory. The tracking feature of gaseous detectors helps suppress
  the background level, resulting in the improvement of the detection
  sensitivity. We study a method based on the convolutional neural
  networks to discriminate double beta decay signals against the
  background from high energy gammas generated by $^{214}$Bi and
  $^{208}$Tl decays based on detailed Monte Carlo simulation. Using
  the 2-dimensional projections of recorded tracks on two planes, the
  method successfully suppresses the background level by a factor
  larger than 100 with a high signal efficiency.  An improvement of
  $62\%$ on the efficiency ratio of $\epsilon_{s}/\sqrt{\epsilon_{b}}$
  is achieved in comparison with the baseline in the PandaX-III
  conceptual design report.}


PACS number(s): 14.60.Pq, 23.40.-s, 07.05.Mh
\section{Introduction}

The Dirac or Majorana nature of neutrinos is one of the most
fundamental questions in particle physics. The conservation of the lepton
number will be violated if neutrinos are Majorana fermions, which is
intricately related to the matter-antimatter asymmetry in our
universe~\cite{Avignone:2007fu}. The so-called neutrinoless double
beta decay (NLDBD) process, in which a nucleus with even atomic number
$Z$ and even neutron number $N$ emits two electrons simultaneously
without neutrinos, is possible if neutrinos are Majorana. Therefore
experimental identification of such rare process would be an important
breakthrough of particle physics.  $^{136}$Xe is a widely used
isotope in experiments searching for NLDBD due to its high abundance
in natural xenon and relatively low cost of enrichment. Experiments using
this target include KamLAND-Zen~\cite{KamLAND-Zen:2016pfg},
EXO-200~\cite{Albert:2014awa}, and NEXT~\cite{Alvarez:2012flf}.

The PandaX-III project plans to construct a ton-scale NLDBD experiment
in the China Jin-Ping underground Laboratory (CJPL) using time
projection chambers (TPCs) filled with high pressure xenon gas with
enriched $^{136}$Xe~\cite{Chen:2016qcd}. The first detector will have
200~kg xenon gas at 10~bar pressure and read out ionized electrons
directly with Micromegas (Micro-MEsh Gaseous Structure)
detectors. Trimethylamine (TMA) mixed with xenon converts part
of the scintillation in xenon to ionization and improves energy
resolution of the TPC.  The energy resolution at $^{136}$Xe
NLDBD Q-value (2.458~MeV) is expected to reach 3\%
(Full-Width-Half-Maximum, FWHM).  The admixture also suppresses
diffusion of ionized electrons and improves the tracking capability.

The main backgrounds in the Region of Interest (ROI) of PandaX-III are
high energy gamma rays from the decay of $^{214}$Bi (2.447~MeV) and
$^{208}$Tl (2.614~MeV) in the detector materials. These background
events may fall in the same energy window of NLDBD Q-Value and mimic a
signal event.  The high pressure gaseous TPC is capable of recording
the tracks of ionizing particles within it, providing additional
information about the events besides the total energy depositions.
Electrons with energy of 1~MeV travel in average 10~cm in the
PandaX-III detector.  At the end of an electron trajectory, the energy
loss per unit length ($dE/dx$) increases dramatically, known as Bragg
peak. NLDBD will have two simultaneous electron tracks in different
directions and thus two distinctive Bragg peaks.  This feature can be
used to distinguish the NLDBD signal from gamma backgrounds. The
topological signatures of $^{136}$Xe NLDBD events and gamma rays have
been studied by the NEXT collaboration~\cite{Cebrian:2013mza,
  Ferrario:2015kta}, and an extra background rejection factor is
achieved by reconstructing the tracks and identifying the end point
energies. It is expected that similar method could provide a
background rejection factor of 35, which serves as a baseline in
PandaX-III.

The topological particle identification method, like other traditional
event classification methods used in particle physics, requires the
reconstruction of designed expert features of the events with
sophisticated algorithms for signal classification. The usage of deep
neural networks in the particle physics for classification without the
aid of designed features has been explored in recent years, and
outstanding performance has been obtained\cite{Baldi:2016fql,
  Barnard:2016qma}.  Convolutional neural networks (CNN), an
artificial neural network originally developed for image analysis,
have gained popularity in particle physics experiments these
years~\cite{Aurisano:2016jvx, Komiske:2016rsd, Madrazo:2017qgh,
  Haake:2017dpr,Luo:2017ncs}.  CNN is especially suited for
signal-background discrimination of electron tracks in the gaseous
TPC. The NEXT collaboration has studied the background rejection power
of CNN in the search of NLDBD by using images with all three
dimensional projections of the tracks, and obtained an improvement
compared with the method based on the same topological
information~\cite{Renner:2016trj}. But such a method cannot be used in
the first phase of PandaX-III directly due to its strip
readout\footnote{See Appendix \ref{sec:limit_strip}}. An alternative
way of data preparation should be considered.

In this work, we apply the CNN technique to the signal and background
discrimination in the PandaX-III experiment, based on a detailed Monte
Carlo (MC) simulation with detector geometry and realistic drifting
of ionized electrons. This article is organized as follows. In
Sec.~\ref{sec:pandax_iii}, we give a brief introduction to the
PandaX-III detector and the properties of NLDBD events in it. In
Sec.~\ref{sec:simulation}, we describe the simulation of NLDBD events
in PandaX-III. Then we introduce the method of classification of NLDBD
signals and backgrounds with convolutional neural networks in
Sec.~\ref{sec:classification}. We present the results based on Monte
Carlo simulation in Sec.~\ref{sec:result}. A short summary is given in
the last section.

\section{The PandaX-III detector and background suppression}
\label{sec:pandax_iii}

A detailed introduction to the PandaX-III detector can be found in the
conceptual design report (CDR)~\cite{Chen:2016qcd}. A high pressure gaseous
TPC is adopted due to its higher energy resolution in comparison with
liquid detectors, and the capability of imaging the electron
tracks. Deposited energy inside the TPC will be released in the form
of scintillation light and ionized electrons. The electrons will drift
towards the ends of the TPC due to the strong electric field and be
finally collected by the readout planes.  The first TPC module will be
filled with 200~kg $90\%$ enriched $^{136}$Xe with $1\%$ TMA
mixture. In the second phase, the ton-scale experiment will have 5
independent detector modules, each of which will contain 200~kg of
xenon gas.

The mixed gas of 10~bar pressure will be enclosed in an Oxygen-Free
High Conductivity (OFHC) copper pressure vessel of cylindrical shape
with a length about 2 m and diameter about 1.5 m. The total volume of
the vessel is about 3.5 m$^3$. A cylindrical TPC, with two drift
regions separated by a cathode plane in the middle, will be placed
inside the vessel. Each drift region will have a drift length of about
1~m and a design drift field of 1000~V/cm, shaped by the field
cage. Two options to build the cylindrical field cage have been
considered. The classical design with copper rings has been used
widely in other experiments, including the PandaX-I~\cite{Cao:2014jsa}
and PandaX-II~\cite{Tan:2016diz} dark matter experiments.  This option
will be the baseline design of PandaX-III and it has been used in all
our MC simulations.  Another design using Diamond-Like-Carbon (DLC)
coated on Kapton films may reduce difficulties during the construction
and overall background rate.  This approach is currently under active
R\&D within PandaX-III collaboration.

A special realization of Micromegas, called Microbulk, will be used in
the first phase of PandaX-III to detect the ionization electrons. An
excellent energy resolution of $3\%$ Full-Width-Half-Maximum (FWHM) at
the NLDBD Q-value is expected based on R\&D results at 10 bar of Xe
with TMA by the T-REX project~\cite{Irastorza:2015dcb}. Each readout
plane of the PandaX-III detector will be covered by 41 specially
designed $20\times20$ cm$^2$ Microbulk Micromegas (MM) modules, and
each module will be read out by 128 channels of X-Y strips of 3 mm
pitch (64 in each direction). The total number of readout channels is
about 10000.

According to the detailed
Geant4~\cite{Agostinelli:2002hh,Allison:2006ve} based MC simulation,
the main background contributions come from the high energy gamma rays
generated by the radioactive descendants of $^{238}$U and $^{232}$Th
inside the detector~\cite{Chen:2016qcd}. Steel bolts and MM modules
contribute to the majority of the backgrounds within the energy window of
$(Q_{\beta\beta}-2\sigma, Q_{\beta\beta}+2\sigma)$, where $\sigma$ is
the corresponding standard deviation at the expected detector
resolution of $3\%$. By assuming the input material activity and
taking the electron diffusion and the detector response into
consideration, the background index (BI) is about $3\times10^{-3}$
counts/(keV$\cdot$kg$\cdot$year)~\cite{Chen:2016qcd}. That means that
about 70 background events would be observed in the energy windows
each year, which is too high for a NLDBD experiment.

Topological information of an event track, such as track length,
shape, and energy deposition per unit distance, can help suppress
background events further. This powerful background suppression
capability has been demonstrated by other
studies~\cite{Irastorza:2015dcb, Ferrario:2015kta}. The two high
energy electrons produced in the NLDBD process will generate large
amount of ionization electrons along their path inside the high
pressure xenon gas and lose energy quickly, generally resulting in
tracks with length of about 15~cm and higher energy depositions at the
ends due to the Bragg peaks of electrons. The gamma background loses
energy mostly through Compton scattering or photoelectric
interactions, and the number of tracks may vary. Ideally, such
topological information can be easily used if the track is fully
reconstructed in 3D spaces. In the first phase of PandaX-III, the $z$
position of the energy deposition, characterizing the drifting time of
electrons, can be extracted easily. But the $x$-$y$ position on the
readout plane can not be determined exactly because of the ambiguity
introduced by multiple strip signals at the same time. Though the 3D
reconstruction is difficult, the projection of the tracks in the $x$-$z$
and $y$-$z$ will be easily acquired. It is necessary to find out a
method to make use of all the information from this incomplete
tracking in order to better discriminate signal and background.

\section{Simulation of NLDBD events in PandaX-III}
\label{sec:simulation}

\begin{figure}[hbt]
  \centering
  \includegraphics[width=0.6\textwidth]{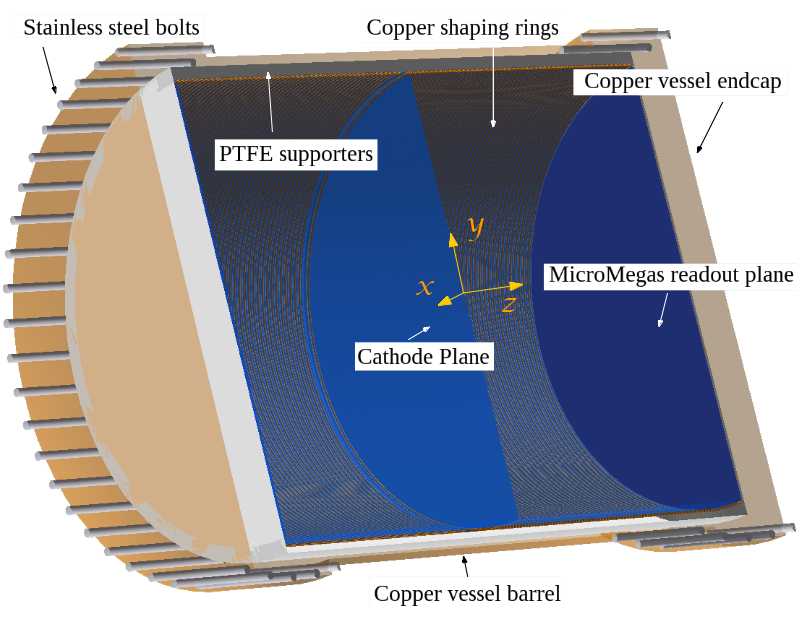}
  \caption{A cross-sectional view of the simulated PandaX-III detector.}
  \label{fig:mc_simulation}
\end{figure}

The study was carried out with MC simulation events. BambooMC, a MC
program based on the Geant4 toolkit, was used to perform detailed
simulation of tracking of electrons and gamma within the PandaX-III
detector. The detailed structure of the detector module, including the
copper vessel, the gas mixture, the field cage with copper rings and
the readout planes, was constructed in the simulation(see
Fig.~\ref{fig:mc_simulation}). The gas mixture contains $99\%$ (mass
fraction) $^{136}$Xe enriched (at the level of $90\%$) xenon gas and
$1\%$ TMA. The electric field of 1000 V/cm along the $z$ direction was
also applied in the simulation.

The DECAY0 package~\cite{Ponkratenko:2000um} was used to generate the
NLDBD signal events, each contains two electrons with a total kinetic
energy around the Q-value. These events were sampled uniformly inside
the gas mixture within the TPC. For the backgrounds, gamma with the
energy of 2447.7 keV (from $^{214}$Bi) and 2614 keV (from $^{208}$Tl)
are sampled from the copper vessel volume with an isotropic angular
distribution.

After the simulation, the information of particle energy depositions
inside the TPC, including the position, time, energy and particle
type, were recorded in the output file, serving as the input of the
following digitization step. Each readout channel's waveform,
expressed by a time series of electrons arriving the corresponding
area in the readout plane, was generated by simulating the generation,
drifting and diffusion of the ionization electrons during the
digitization.

For each energy deposition, the number of ionization electrons is
randomly sampled with the W-value of 21.9 eV and Fano factor of
0.14~\cite{Aprile:2009dv}. The drifting velocity and diffusion
parameters of ionization electrons were calculated with the
Magboltz~\cite{magboltz} package through the interface of
Garfield++~\cite{garfield} by taking the constituent of the gas
mixture and electric field into consideration. In this study, we
continued to use the drifting velocity of 1.87 mm/$\mu$s, the
transverse diffusion parameter of $1.02\times10^{-2}$ cm$^{1/2}$ and
the longitudinal diffusion parameter of $1.39\times10^{-2}$
cm$^{1/2}$~\cite{Chen:2016qcd}.

For each ionization electron, the time and position when it arrives
the readout plane is calculated with the initial position of the
energy deposition together with the parameters. Then the possible
fired readout channel is determined according to the arrangement of
the MM modules and channels. The electron loss during the drifting is
ignored. In real experiment, the gas will be purified continuously by
the online recycling system.

\begin{figure}[hbt]
  \centering
  \includegraphics[width=0.6\textwidth]{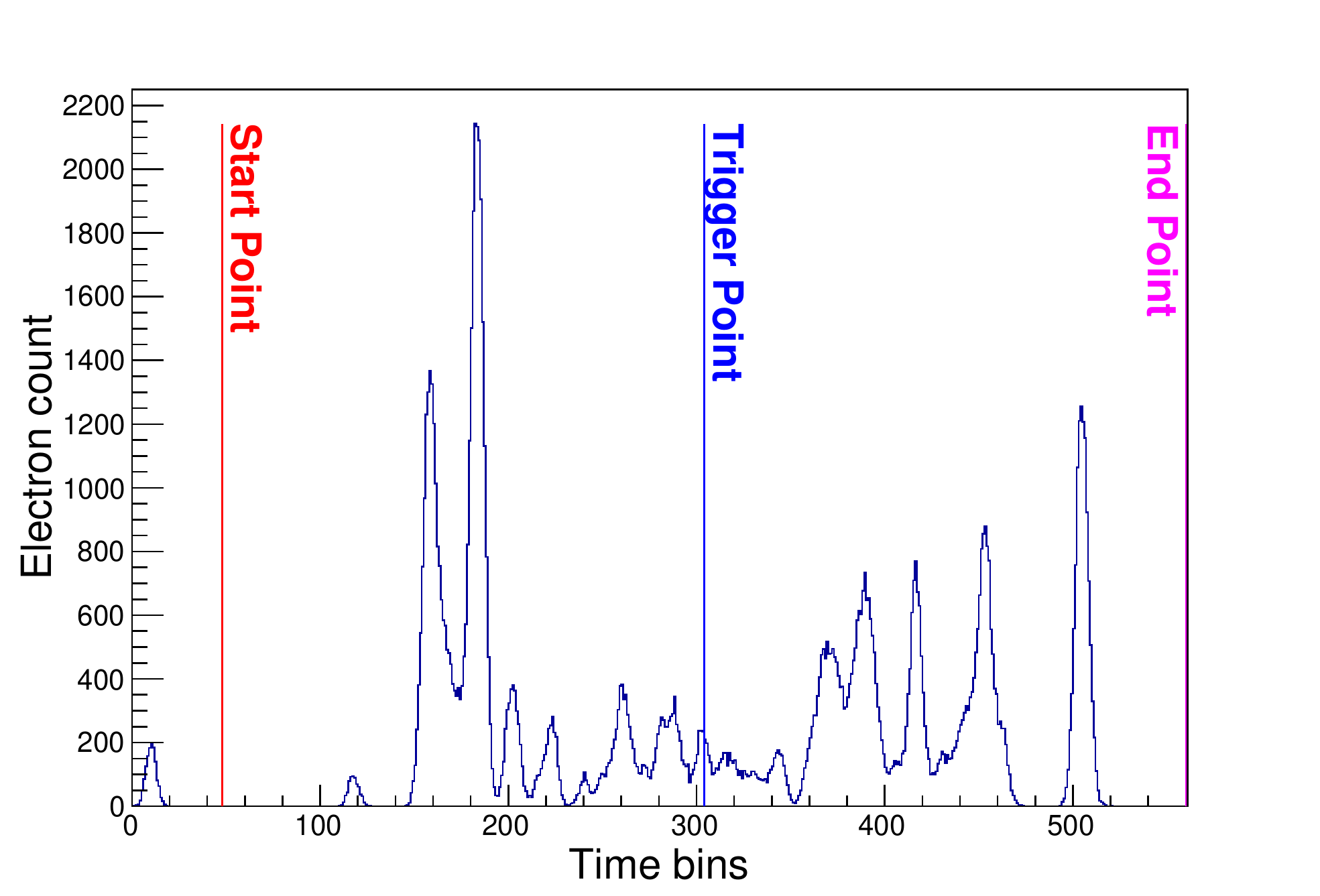}
  \caption{The number of electrons hitting a readout plane in an event by time. The bin width is 20ns. Time 0 is defined as the time when the first electron reaches the readout plane. The time series between the start point and the end point are finally recorded.}
  \label{fig:channel_time_series}
\end{figure}

We obtained a set of time series of recorded electron number for all
the readout channels after iterating all the depositions in the
TPC. The width of the time bin is determined assuming a sampling rate
of 5 Ms/sec. An example of such a time series is shown in
Fig.~\ref{fig:channel_time_series}. Because the readout window is
limited to 102.4 $\mu$s (512 time bins) and the maximum drifting time
in one chamber of the PandaX-III TPC is about 535 $\mu$s, the
waveforms may contain only part of the long time series. To simulate
the trigger, We used the integrated energy, which is translated from
the number of electrons, in a sliding window of 256 bins along the
time axis, as a trigger. The event is ``triggered'' when the energy
exceeds $Q_{\beta\beta}/2=1.229$ MeV. Two hundred fifty-six bins
before the trigger signal and 256 bins after are rept in the final
waveform. These waveforms simulate the outputs from real detector
electronics. The summation of the total number of electrons are
translated directly into the readout energy without additional
smearing applied.

After the detailed simulation and digitization, we found the detection
efficiency of NLDBD signal is 56.2$\%$ within the energy window of
$Q_{\beta\beta}\pm2\sigma$, i.e., $[2395, 2520]$~keV, assuming a
detector resolution of $3\%$ full width half maximum (FWHM). The value
is consistent with that reported in the PandaX-III CDR.

\section{Event Classification with CNN}
\label{sec:classification}

The rapid development and application of the deep neural networks,
especially the usage of CNN in image classification in recent years
~\cite{NIPS2012_4824,DBLP:journals/corr/ZeilerF13}, provide a
possibility for the discrimination of signal and background in
PandaX-III without fully reconstructing the tracks.

\subsection{Introduction of CNN}
\label{sec:cnn_brief}

A detailed introduction to the CNN can be found in
Ref.~\cite{DBLP:journals/corr/ZeilerF13}. We introduce the basic ideas
here. Similar to the ordinary neural networks, the basic block of CNNs
is neuron, which generates output value according to the input and its
own parameters. Neurons are grouped into layers by their different
functionalities, and the network is a stacking of the layers. Each layer
of the network can be regarded as a non linear tranformation that maps
a tensor(multidimensional arrays) to another tensor, so the whole network
is mostly like a complicated transformation which tries to fit the input
tensors to the output tensors.

CNNs are specially designed to work with image data, by reducing the
parameter space with ``convolution''. The first layer of a CNN reads
data from images of certain size, and the last layer generates the
result of classification by arranging all neurons in a row.  A CNN
may contain one or more convolutional layers, in which each neuron is
used to compute the dot product ({\em convolution}) of its weight and
values from a small region in the input volume. With such structures,
CNNs are capable of learning features from given images during the
training procedure so that they can be used to classify new images by
recognizing their features with the same algorithm.

It was shown that significant improvement on the performance had been
achieved with higher network
depth~\cite{DBLP:journals/corr/SimonyanZ14a}. But a degradation
problem appears when the network becomes deeper, together with higher
training error~\cite{DBLP:journals/corr/He014}.  The residual network
(ResNet), introduces shortcut connections between nonadjacent layers
to mitigate the problem~\cite{DBLP:journals/corr/HeZRS15} and
shows outstanding performance.
In our study, we used the 50-layer ResNet-50 within the
Keras~\cite{chollet2015keras} deep learning library\footnote{A
  comparison of models we have used can be found in Appendix
  \ref{sec:cnn_comp}.}. The input and output layers of the network have been modified so that it can accept our input data and generate an output
value between 0 (exact background) and 1 (exact signal). The detailed
network structure is given in Table~\ref{tab:structure}.

\begin{table}[hbt]
  \centering
  \begin{tabular}{ccccc}
    \\\hline
layer name & layer type &  output tensor  & layer attribute & repetition \\\hline
input\_1  & InputLayer & 240, 240, 3 &  & \\ \hline
conv1 block & Convolution2D & 120, 120, 64 & 7$\times$7, 64 &  \\ \hline
pooling & MaxPooling2D & 59, 59, 64 &  &  \\ \hline
\multirow{3}{*}{conv2 block} & Convolution2D & 59, 59, 64 & 1$\times$1, 64 & \multirow{3}{*}3\\
 & Convolution2D & 59, 59, 64 & 3$\times$3, 64 & \\
 & Convolution2D & 59, 59, 256 & 1$\times$1, 256 & \\ \hline
\multirow{3}{*}{conv3 block} & Convolution2D & 30, 30, 128 & 1$\times$1, 128 & \multirow{3}{*}{4} \\
 & Convolution2D & 30, 30, 128 & 3$\times$3, 128 & \\
 & Convolution2D & 30, 30, 512 & 1$\times$1, 512 & \\ \hline
\multirow{3}{*}{conv4 block} & Convolution2D & 15, 15, 256 & 1$\times$1, 256 & \multirow{3}{*}{6} \\
 & Convolution2D & 15, 15, 256 & 3$\times$3, 256 & \\
 & Convolution2D & 15, 15, 1024 & 1$\times$1, 1024 & \\ \hline
\multirow{3}{*}{conv5 block} &  Convolution2D & 8, 8, 512 & 1$\times$1, 512 & \multirow{3}{*}{3}\\
 & Convolution2D & 8, 8, 512 & 3$\times$3, 512 & \\
 & Convolution2D & 8, 8, 2048 & 1$\times$1, 2048 & \\ \hline
pooling & AveragePooling2D & 1, 1, 2048 &  & \\ \hline
flatten & Flatten & 2048 &  & \\ \hline
dense & Dense & 256 & relu & \\  \hline
dropout & Dropout & 256 &  & \\   \hline
dense & Dense & 1 & sigmoid & \\ \hline
  \end{tabular}
  \caption{The structure of the modified ResNet-50. The ``repetition'' column indicates the number of times the block appears in the network, and the default value is 1.}
  \label{tab:structure}
\end{table}

\subsection{Preparation of Input Data}

\begin{figure}[hbt]
\centering
\begin{subfigure}[t]{0.22\textwidth}
  \centering
  \includegraphics[width=\textwidth]{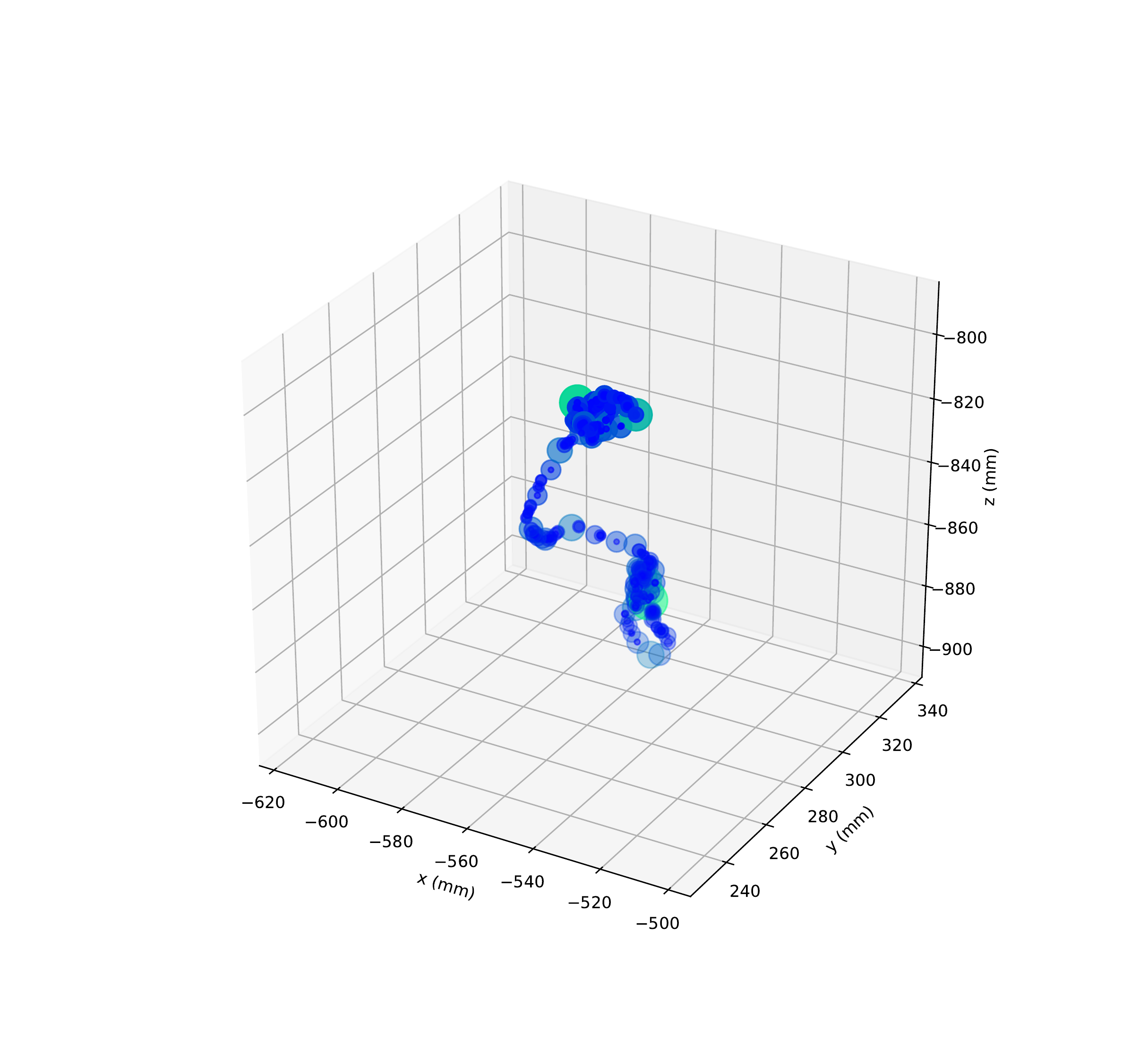}
  \caption{}
\end{subfigure}
\begin{subfigure}[t]{0.44\textwidth}
  \centering
  \includegraphics[width=\textwidth]{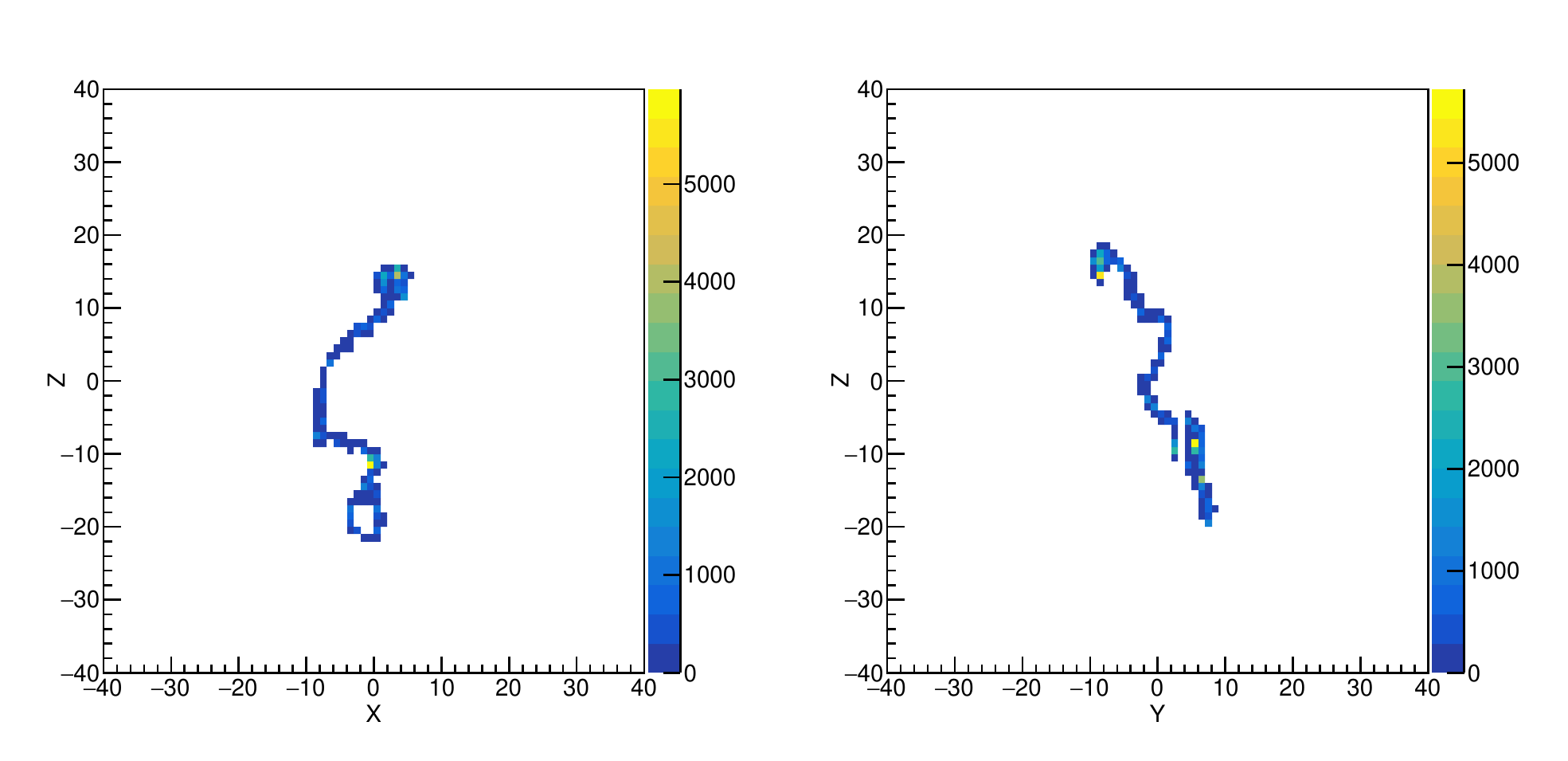}
  \caption{}
\end{subfigure}
\begin{subfigure}[t]{0.16\textwidth}
  \centering
  \setlength{\fboxsep}{0pt}
  \fbox{\includegraphics[width=\textwidth]{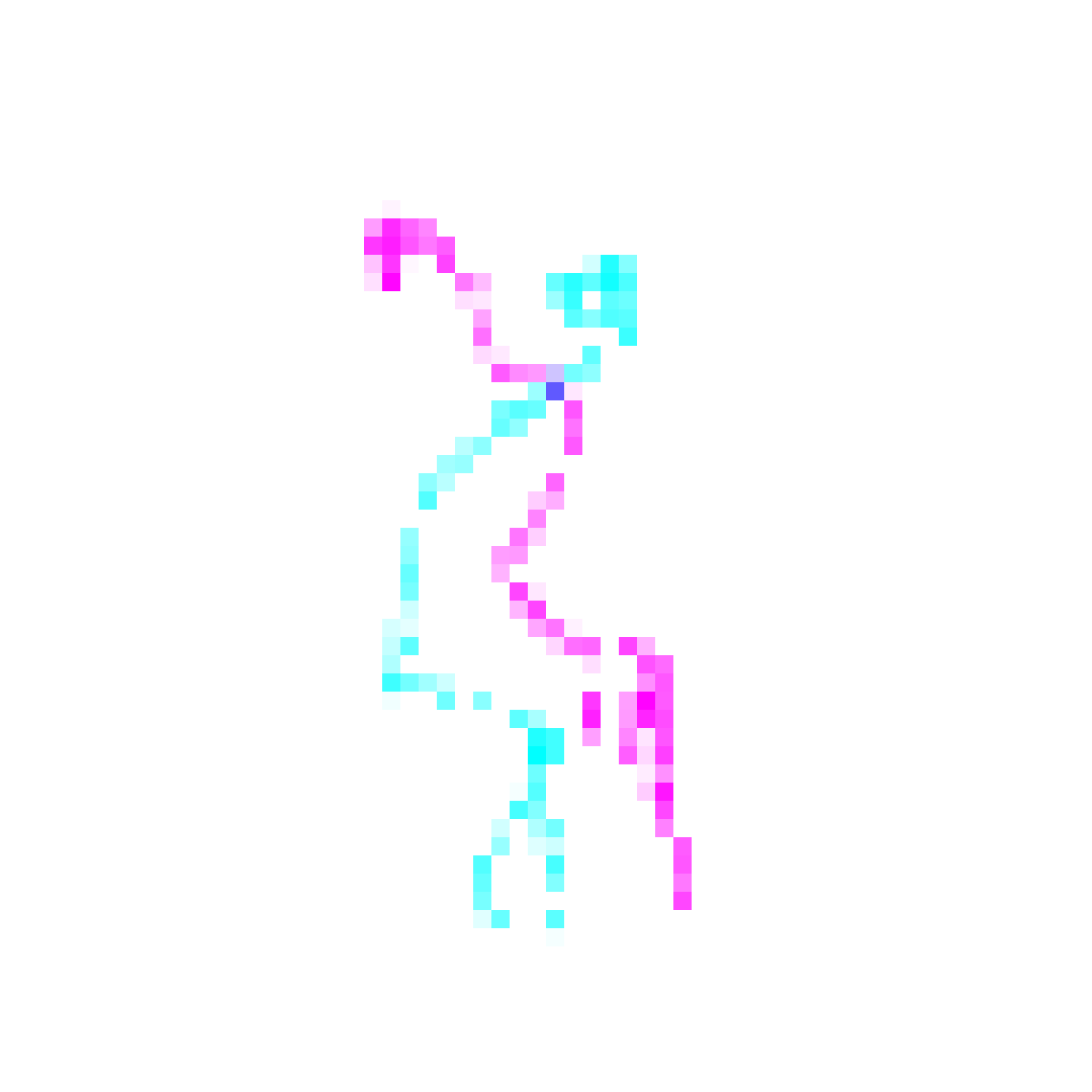}}
  \caption{}
\end{subfigure}

\begin{subfigure}[t]{0.22\textwidth}
  \centering
  \includegraphics[width=\textwidth]{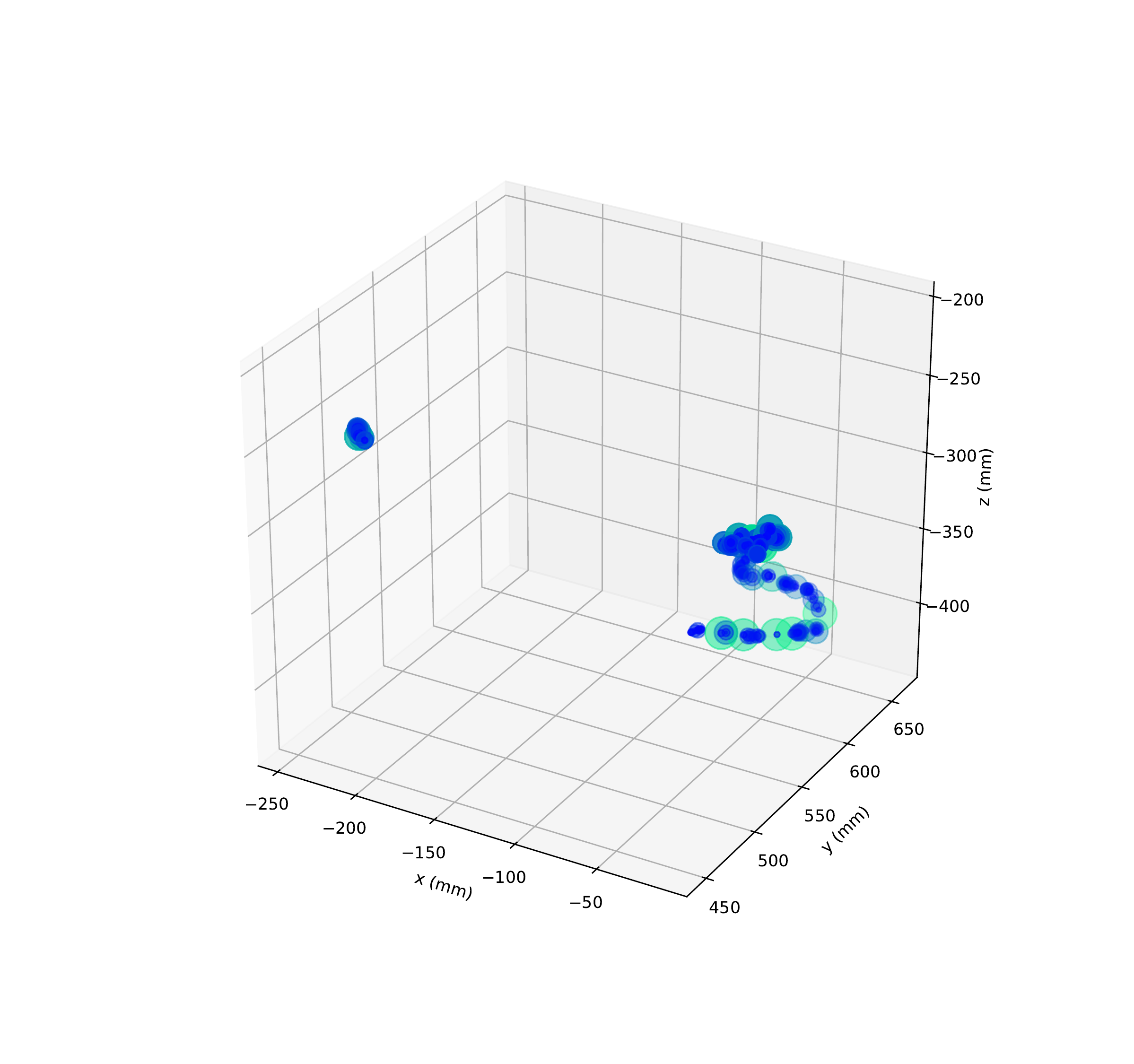}
  \caption{}
\end{subfigure}
\begin{subfigure}[t]{0.44\textwidth}
  \centering
  \includegraphics[width=\textwidth]{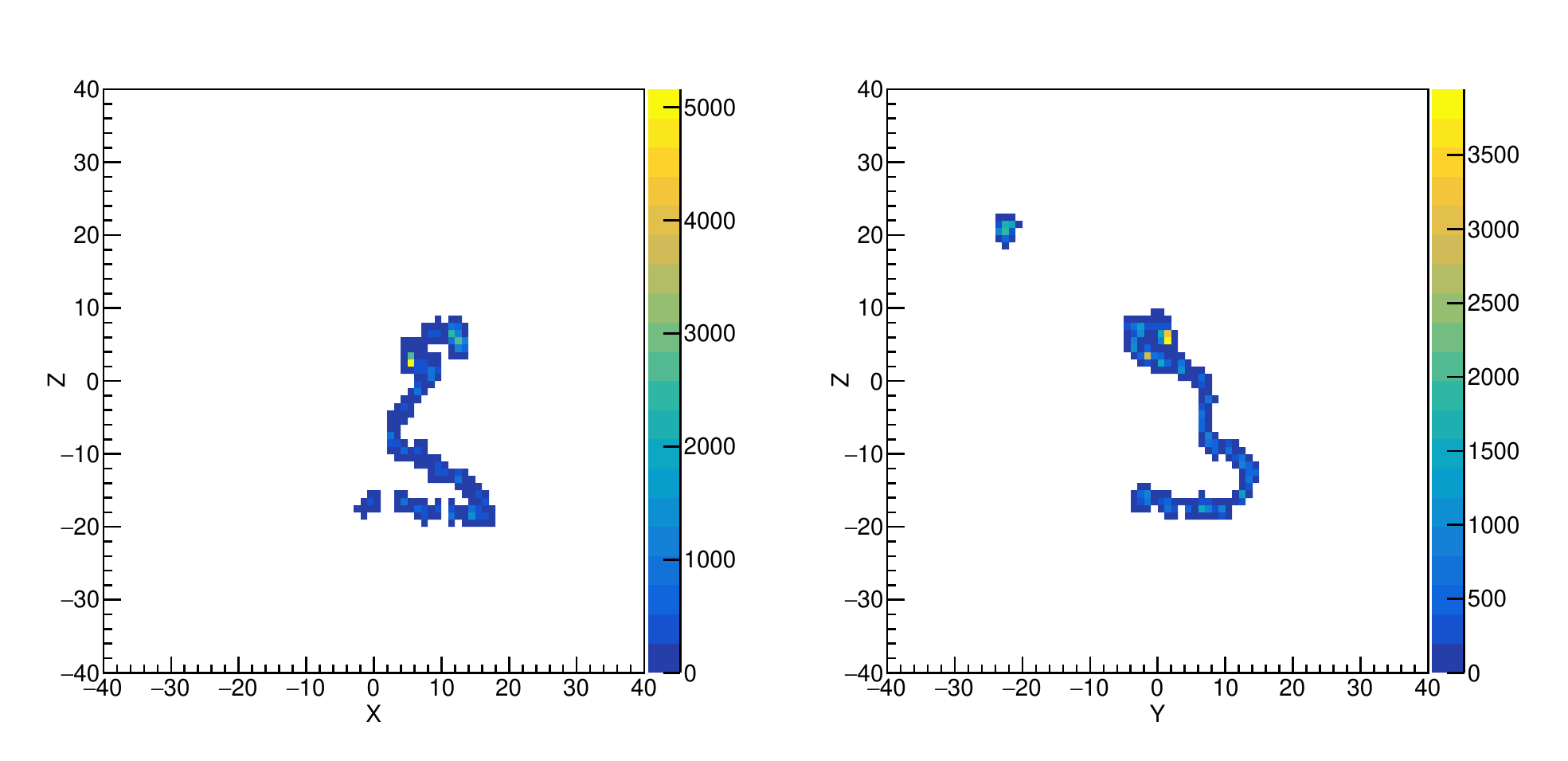}
  \caption{}
\end{subfigure}
\begin{subfigure}[t]{0.16\textwidth}
  \centering
  \setlength{\fboxsep}{0pt}
  \fbox{\includegraphics[width=\textwidth]{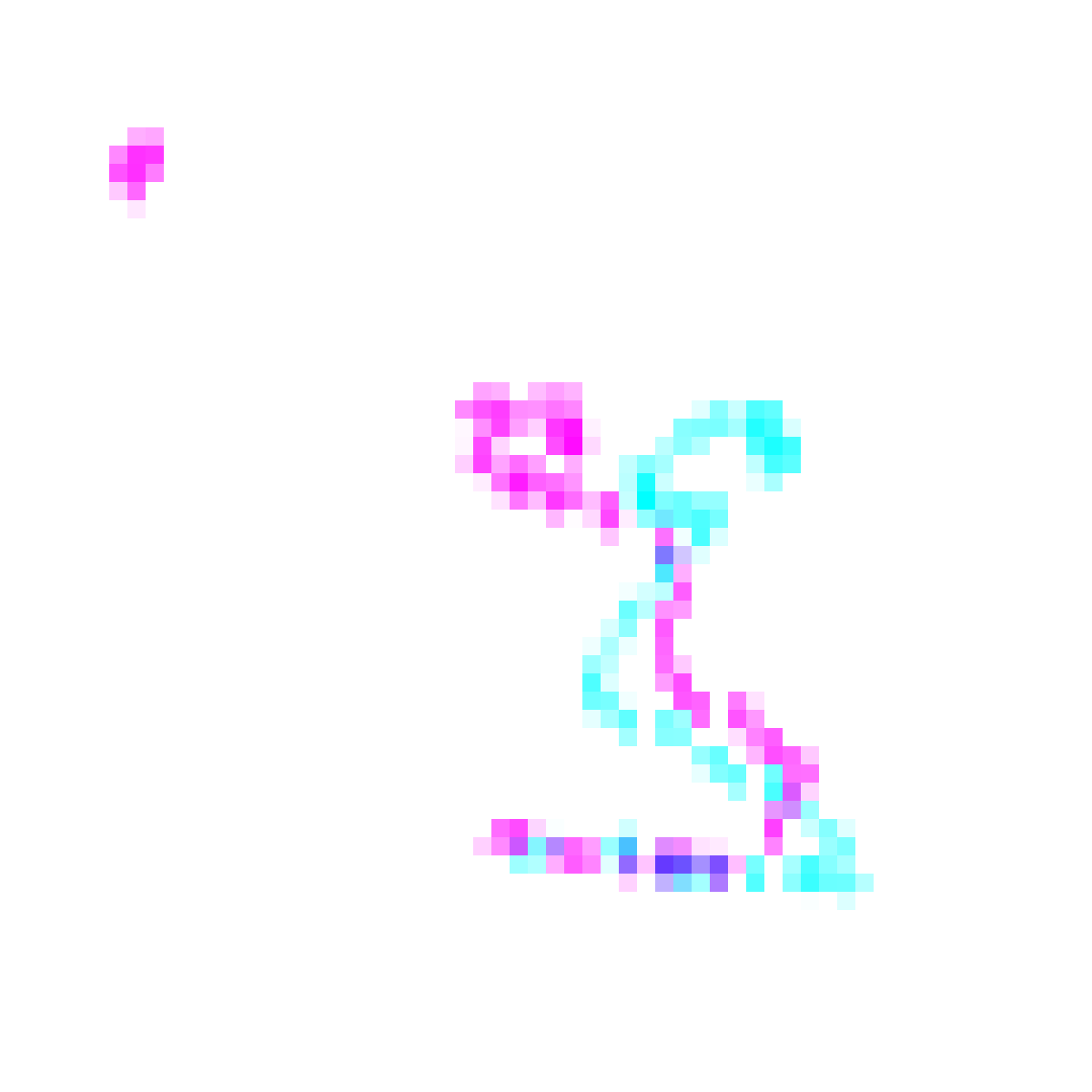}}
  \caption{}
\end{subfigure}

\caption{Examples of mapping from recorded events to images. Top: $^{136}$Xe NLDBD event; Bottom: Gamma background event from $^{214}$Bi. (a) and (d): The raw Monte Carlo 3D hits map. (b) and (e): The $x$-$z$ and $y$-$z$ projection of the event after reconstruction. (c) and (f): The resulting images for training. For better visualization, color inversion and enhancement were applied to the final images. The cyan color stands for the red channel, and the magenta color for the green channel.}
\label{fig:image_mapping}
\end{figure}

Since CNNs need images as the input, the MC events after digitization
need to be converted to images. We converted each event to an image of
$60\times60$ pixels with the PNG format.  For each readout signal by
the vertically/horizontally arranged strips, its coordinate of
$(x, z)$/$(y,z)$ was mapped to the image coordinates $(X, Y)$, and
corresponding energy was encoded in the red/green color channel.  In
this way the two projections were merged into one picture. The center
of the image was chosen to be the energy-weighted center of the hits to
ensure most of the energy of the event could be encoded in the
image. Each pixel in an image represents an area of $3\times3$ mm$^2$
in the $xz$ and $yz$ planes of the corresponding event, so the total
area is $180\times180$ mm$^2$. An example of such a mapping is shown
in Fig.~\ref{fig:image_mapping}.

We generated $5.6\times10^5$ images from simulated NLDBD signal events
and $5.6\times10^5$ images from the high energy gamma background
events. Within these images, $80\%$ (training set) are used for the
training of the CNN, $10\%$ (validation set) are used for the
validation, and remain $10\%$ (testing set) are used for the checking
of the power of discrimination.

\subsection{Results of Event Discrimination}
\label{sec:result}

\begin{figure}[hbt]
  \centering
  \includegraphics[width=0.6\linewidth]{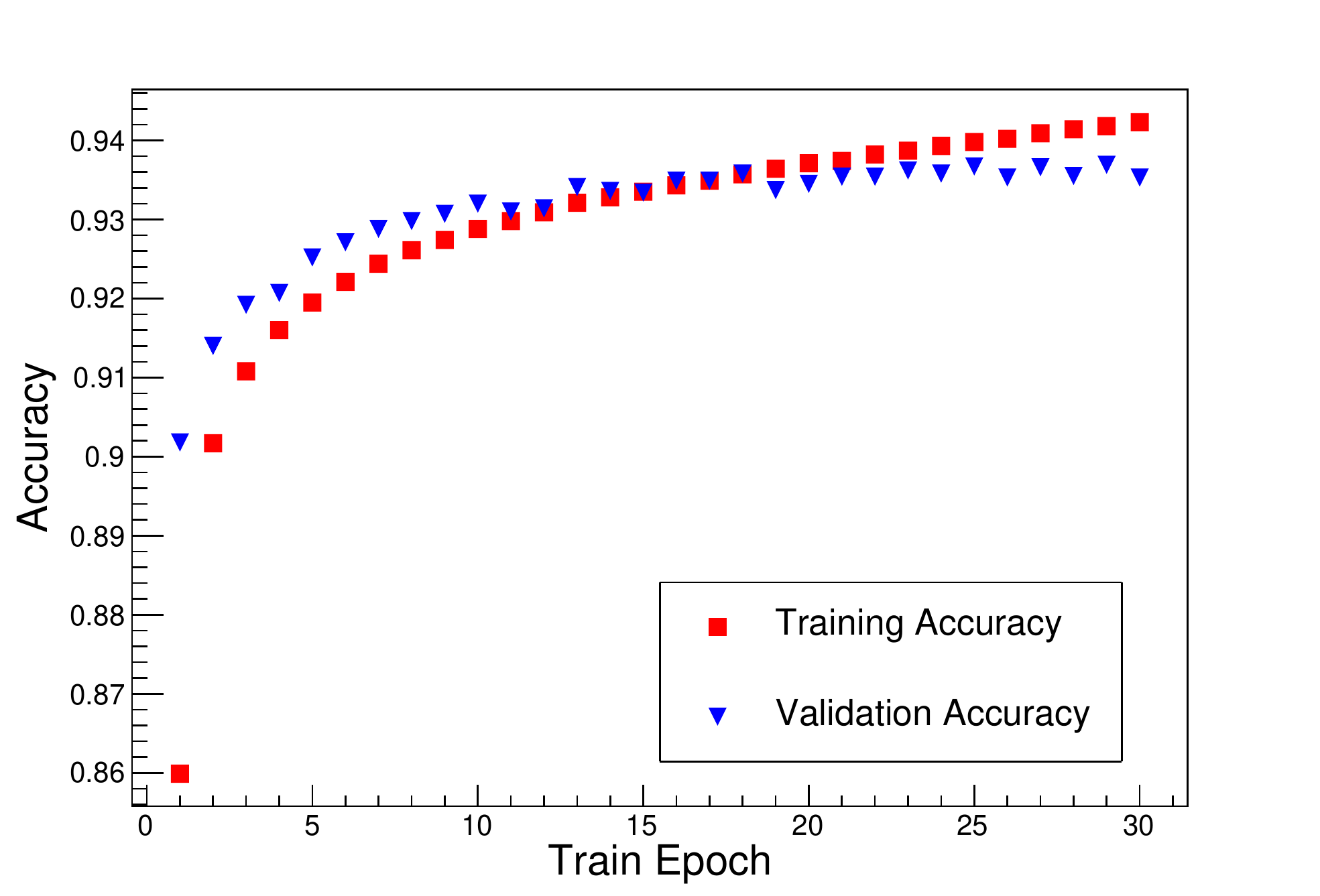}
  \caption{The evolution of the training/validation accuracy with epochs.}
  \label{fig:train_par}
\end{figure}

The modified ResNet-50 was trained with the training set on a
workstation with two NVidia GeForce 1080 GPUs for 30
epochs\footnote{An epoch is a complete pass through a given dataset.}.
To prevent overfitting and use the input data more efficiently,
real-time data augmentation, such as the random operations of
rotation, shifting and shearing, have been applied to the training
images. The network parameters were updated in each epoch. The value
of ``{\em accuracy}'', defined as the ratio between the number of
correctly recognized events over the total number of events with the
 threshold of 0.5, is a measurement of the agreement of model
prediction in comparison with the data. We plot the training and
validation accuracies in Fig.~\ref{fig:train_par}. The network became
overfitting apparently after the 20th epoch due to the validation
accuracy became smaller than the training accuracy, though the
difference is small. The training accuracy and validation accuracy are
nearly identical from the 16th to 20th epoch, and the relative
differences are smaller than $0.1\%$. To avoid the bias introduced by
the arbitrary selection of models, we used all the trained models from
the 16th to 20th epochs in following study.

\begin{figure}
  \centering
  \includegraphics[width=0.6\textwidth]{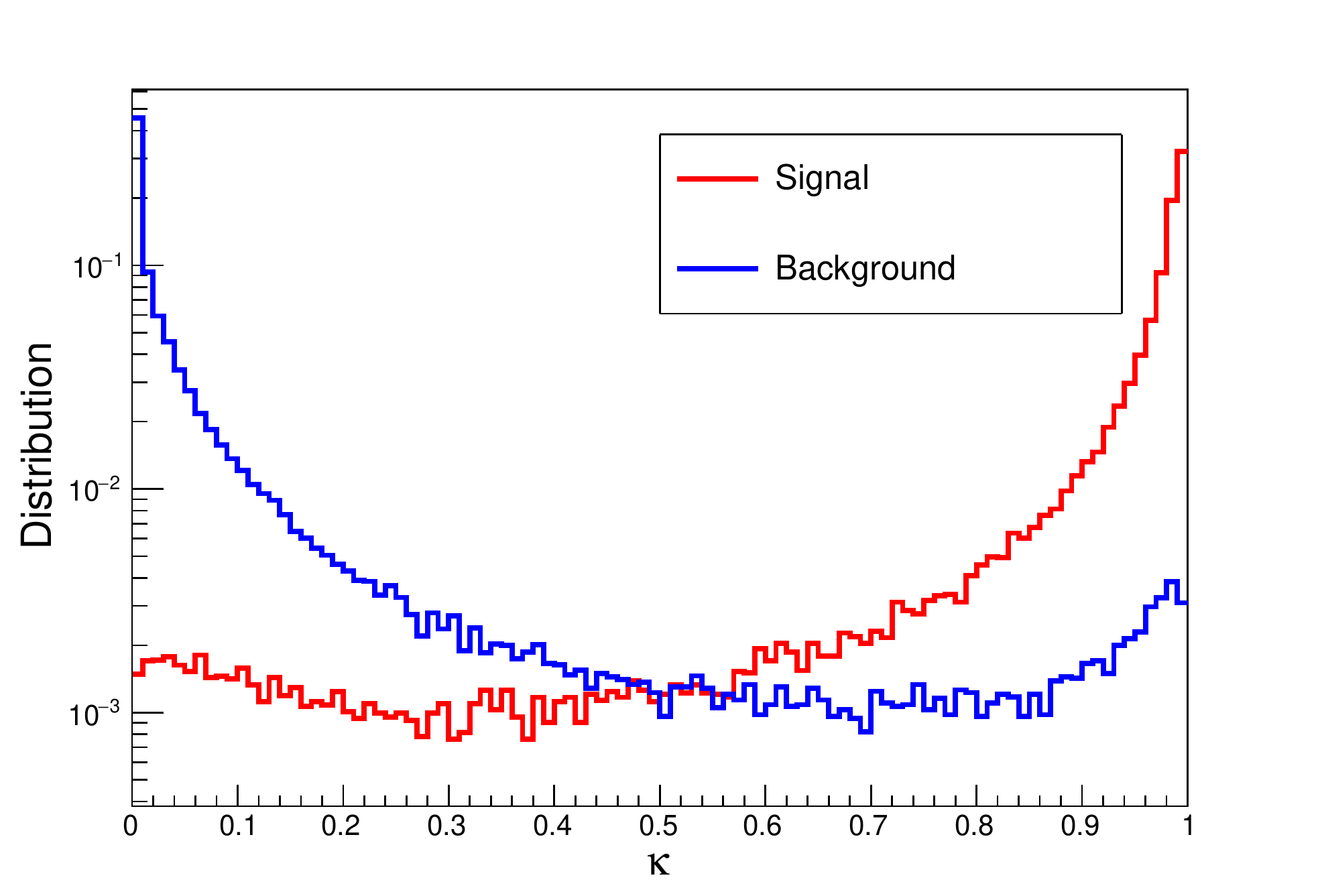}
  \caption{The distributions of $\kappa$ for NLDBD signals (red) and
    high energy gamma backgrounds (blue) from model-16 in testing
    dataset. The rising of $\kappa$ for backgrounds is resulted from
    the signal-like background events.}
  \label{fig:res_dis}
\end{figure}

The trained models were used to classify the images in the testing
set. For each input image, the output value $\kappa$ is a number
between 0 and 1, representing how it looks like a background (0) and a
signal (1). The distributions of $\kappa$ for NLDBD signal and
backgrounds from model-16 (trained model in the 16th epoch) are shown
in Fig.~\ref{fig:res_dis}.

\begin{figure}
  \centering
  \includegraphics[width=0.6\textwidth]{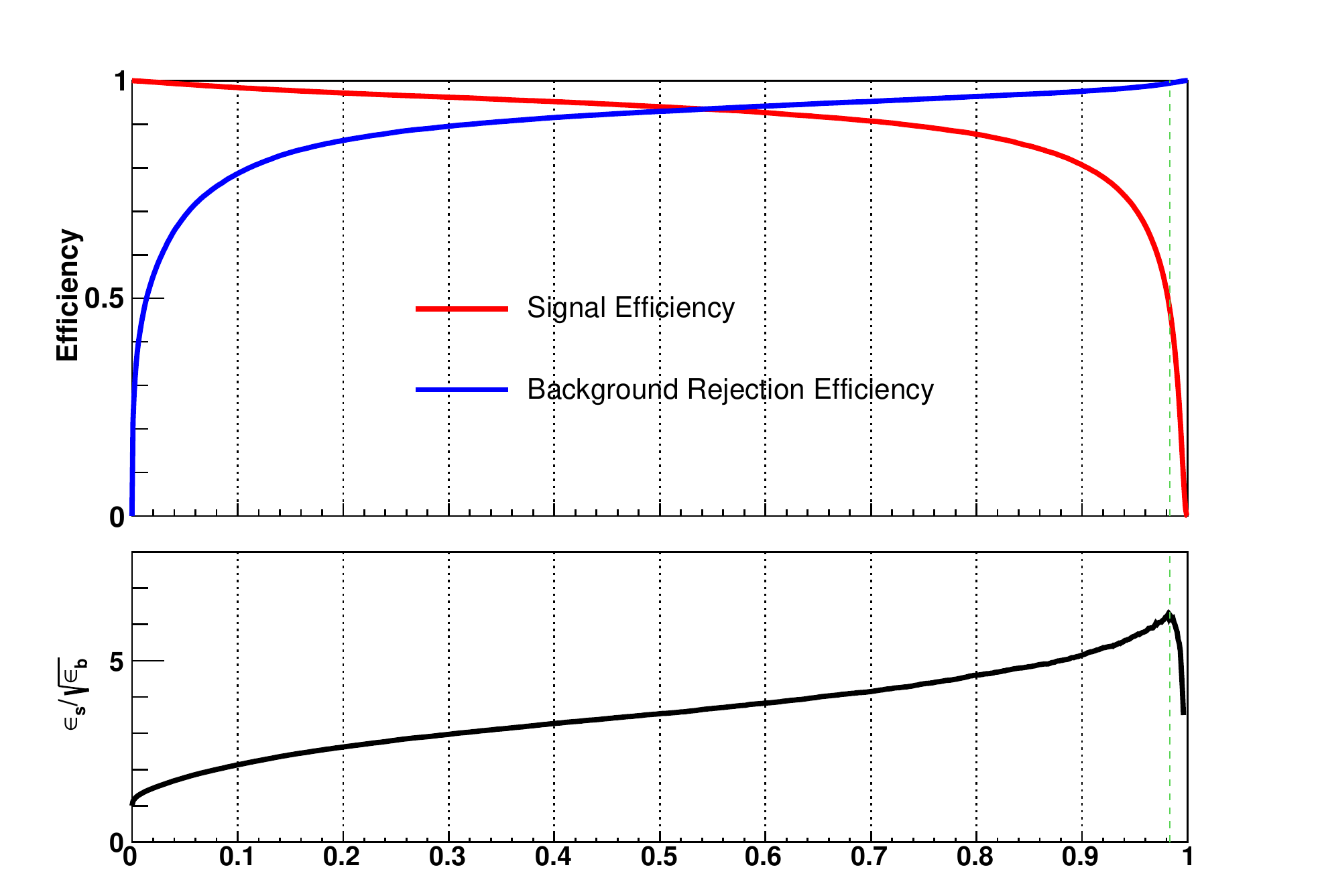}
  \caption{Top: The signal efficiency $\epsilon_{s,cnn}$ (red) and the
    background rejection efficiency $1-\epsilon_{b,cnn}$ (blue) as a
    function of $\kappa_c$ from model-16. Bottom: The efficiency
    ratio $\epsilon_{s,cnn}/\sqrt{\epsilon_{b,cnn}}$ as a function of
    $\kappa_c$ from model-16. The optimized $\kappa_c$ is
    plotted as a green dashed line.}
  \label{fig:eff_dis}
\end{figure}

A cutting threshold can be applied on the distribution to obtain the
signal and background efficiency. The efficiency curves for signal and
background events of model-16 at different cut values are given in
Fig.~\ref{fig:eff_dis}. The selection of the threshold can be
optimized by the definition of figure of merit (FOM). The commonly
defined FOM is proportional to the ratio between the final number of
signal events $s$ and the square root of final number of background
events $b$, or
\begin{equation}
  \label{eq:fom_def}
  \text{FOM} \propto \frac{s}{\sqrt{b}} = \frac{s_d}{\sqrt{b_d}}\cdot \frac{\epsilon_{s,cnn}}{\sqrt{\epsilon_{b,cnn}}} \propto \frac{\epsilon_{s,cnn}}{\sqrt{\epsilon_{b,cnn}}},
\end{equation}
where $\epsilon_{s,cnn}$ and $\epsilon_{b,cnn}$ are the efficiencies
for signal and background of CNN at a given cut $\kappa_c$,
respectively, and $s_d$ and $b_d$ are the number of detected signal
events and backgrounds before the CNN discrimination.

\begin{figure}
  \centering
  \includegraphics[width=0.6\textwidth]{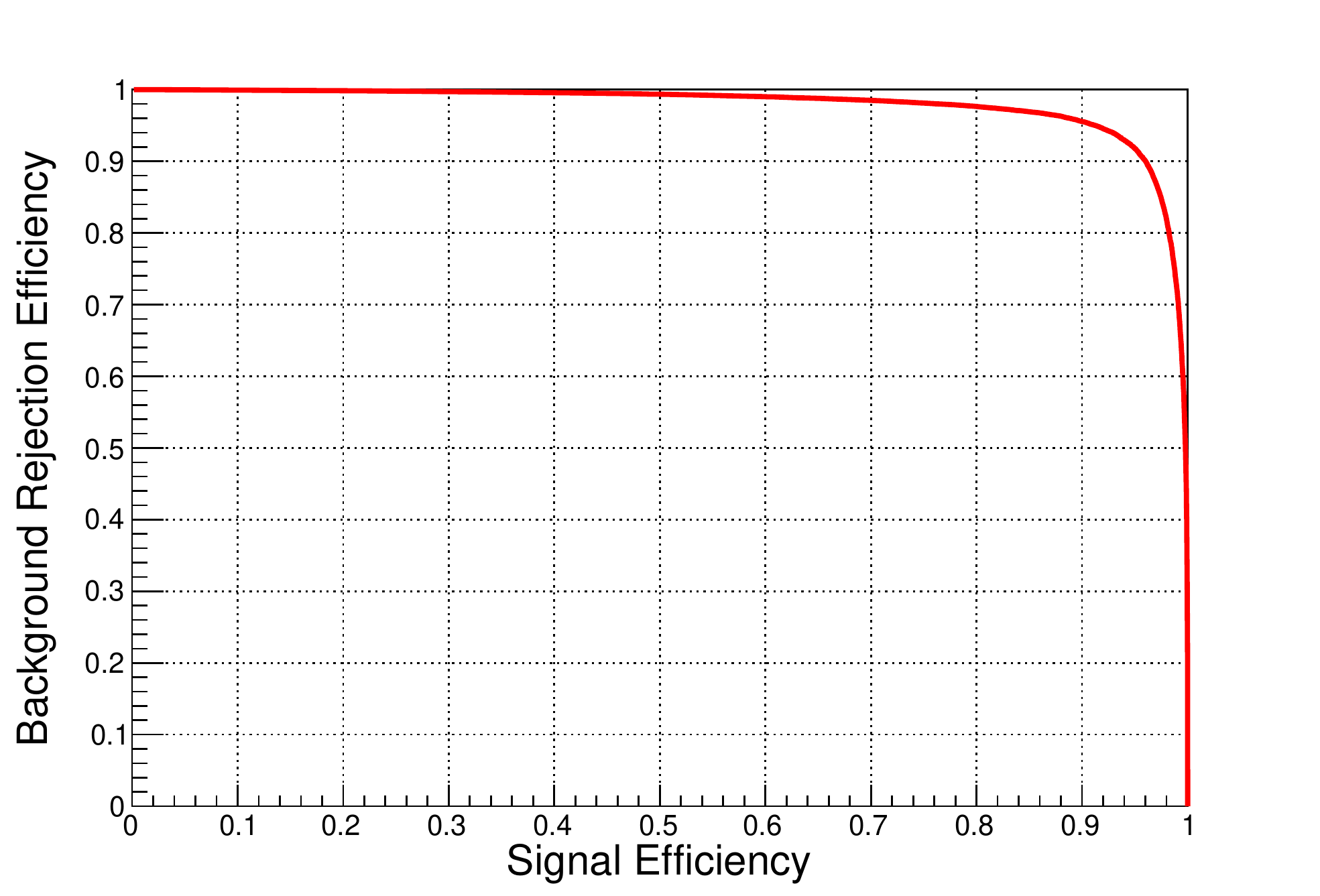}
  \caption{The background rejection efficiency $1-\epsilon_{b,cnn}$
    versus the signal efficiency $\epsilon_{s,cnn}$ from model-16.}
  \label{fig:roc}
\end{figure}

The optimized $\kappa_c$ can be found by maximizing the efficiencies
ratio of $\epsilon_{s,cnn}/\sqrt{\epsilon_{b,cnn}}$. The efficiencies
as functions of $\kappa_c$ from model-16 is plotted in
Fig.~\ref{fig:eff_dis}. The background rejection efficiency versus
signal efficiency is shown in Fig.~\ref{fig:roc}. The corresponding
signal and background efficiencies at optimized $\kappa_c$ in each
model are given in Table.~\ref{tab:efficiencies}.

\begin{table}[hbt]
  \centering
  \begin{tabular}{cccccc}
    \\\hline
    epoch & optimized $\kappa_c$ & $\epsilon_{s,cnn}$ & $ 1-\epsilon_{b,cnn}$ &$\epsilon_{s,cnn}/\sqrt{\epsilon_{b,cnn}}$ & final BI\\\hline
    16 & 0.983 & 0.475 & 0.9943 & 6.264 & $1.775\times10^{-5}$ \\
    17 & 0.976 & 0.569 & 0.9916 & 6.196 & $2.605\times10^{-5}$ \\
    18 & 0.981 & 0.487 & 0.9936 & 6.098 & $1.968\times10^{-5}$ \\
    19 & 0.966 & 0.540 & 0.9923 & 6.165 & $2.369\times10^{-5}$ \\
    20 & 0.976 & 0.520 & 0.9928 & 6.145 & $2.215\times10^{-5}$ \\\hline
    average &  &  &  & $6.174\pm0.055$ \\\hline
  \end{tabular}
  \caption{The optimized $\kappa_c$, corresponding signal efficiency $\epsilon_{s,cnn}$, background rejection efficiency $1 - \epsilon_{b,cnn}$, the ratio of $\epsilon_{s,cnn}/\sqrt{\epsilon_{b,cnn}}$ and final BI (count$\cdot$kg$^{-1}\cdot$keV$^{-1}\cdot$year$^{-1}$). The BI before the CNN discrimination is $3.088\times10^{-3}$ count$\cdot$kg$^{-1}\cdot$keV$^{-1}\cdot$year$^{-1}$~\cite{Chen:2016qcd}.}
  \label{tab:efficiencies}
\end{table}

\begin{figure}[hbt]
  \centering
  \includegraphics[width=0.9\textwidth]{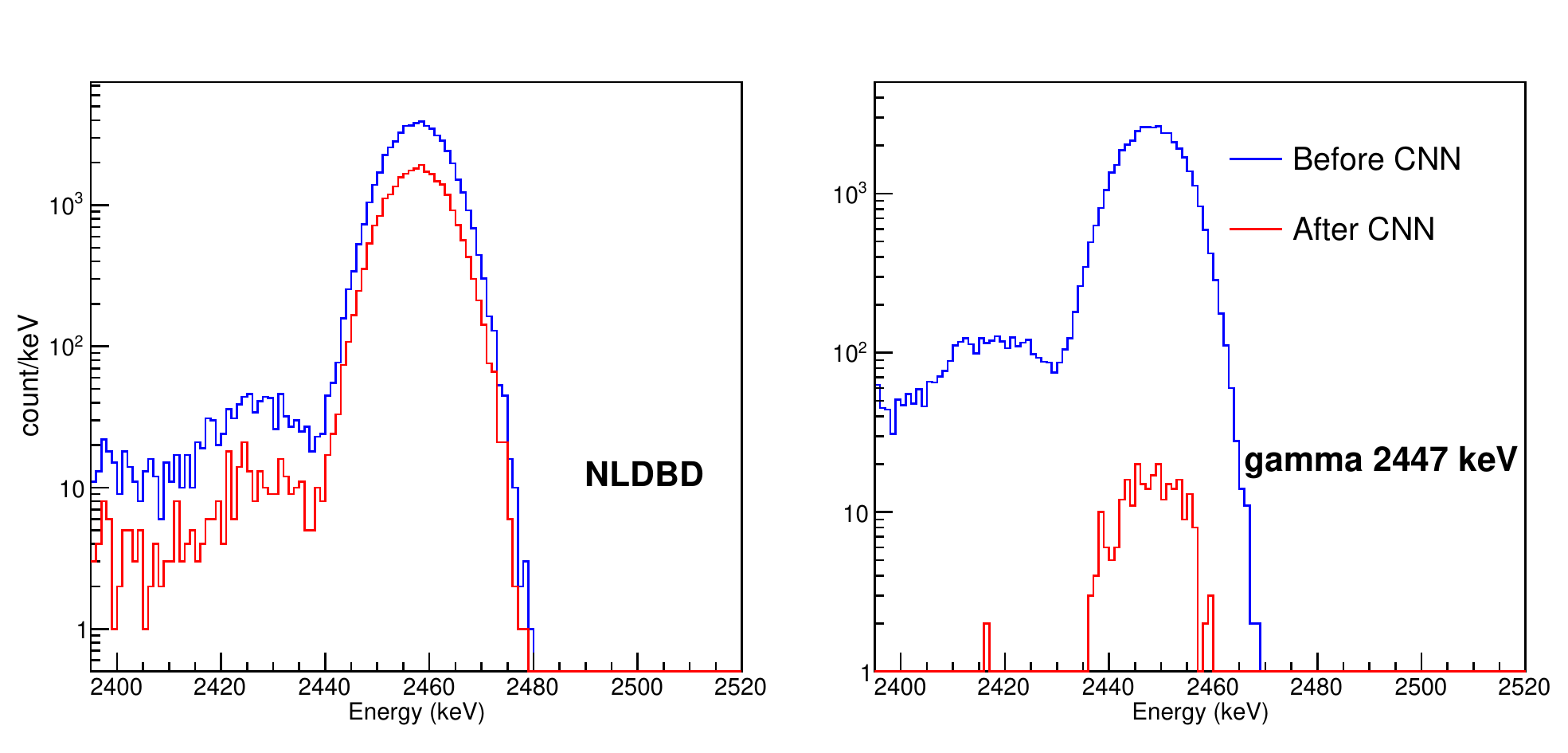}
  \caption{The reconstructed energy spectra of signal and backgrounds before and after the optimal cut $\kappa_c$ from model-16. left: NLDBD signal events; right: background events from $^{214}$Bi. The spectra are not normalized.}
  \label{fig:e_spec}
\end{figure}

We obtained different signal efficiency and background rejection
efficiency with the trained network in different epochs. The final
background index has been suppressed by a factor larger than 100 in all
the models. The relative error of the efficiency ratio
$\epsilon_{s,cnn}/\sqrt{\epsilon_{b,cnn}}$ is only $0.09\%$,
indicating the stability of the discrimination and should be treated
as the systematic error from model selection.
The reconstructed energy spectra of signal and background events
before and after the optimal cut $\kappa_c$ from model-16 are plotted
in Fig.~\ref{fig:e_spec}. The shape of correctly identified signal events
is similar to that before the classification.

\begin{figure}[hbt]
  \centering
\begin{subfigure}[t]{0.44\textwidth}
  \centering
  \includegraphics[width=\textwidth]{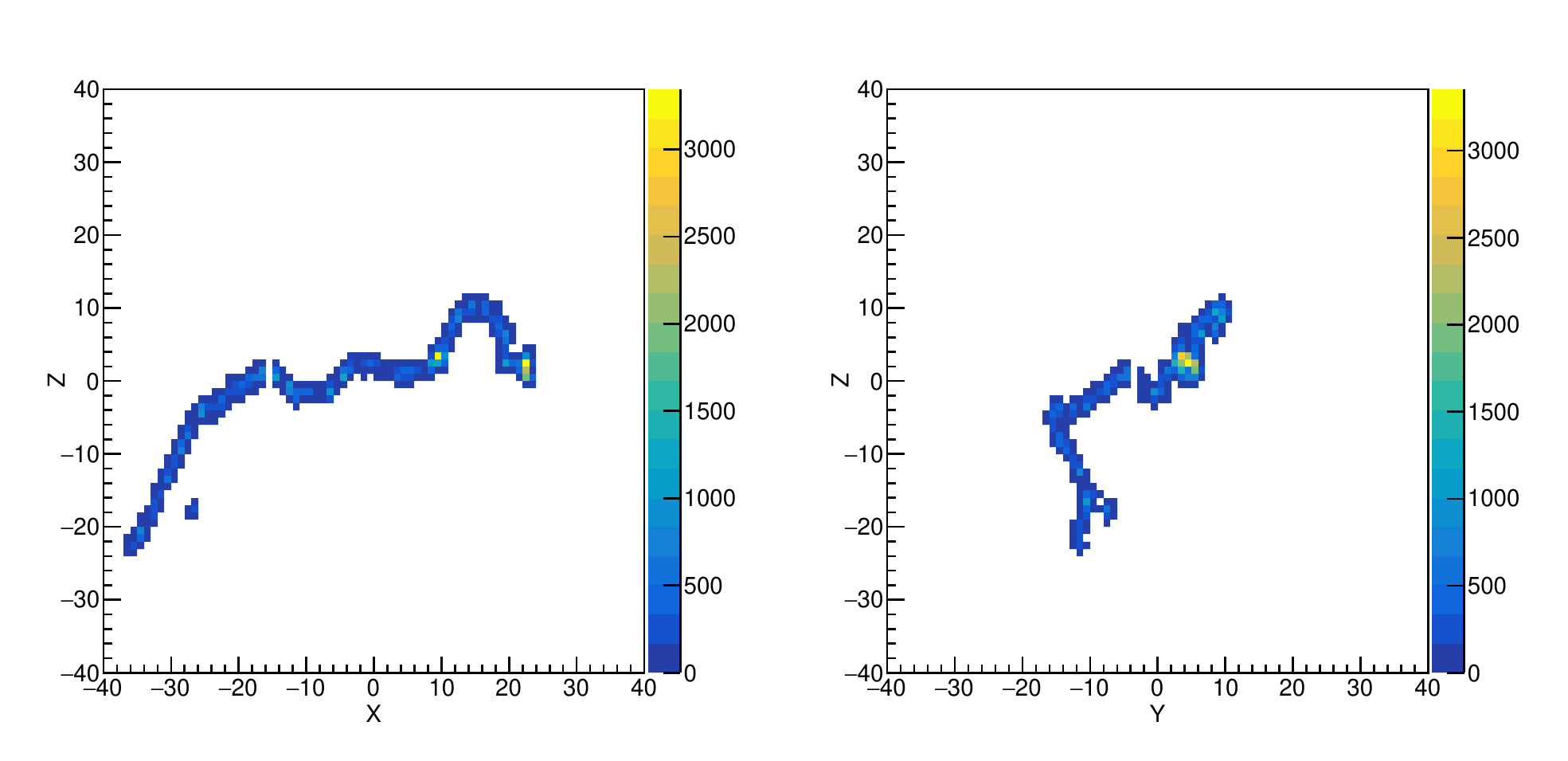}
  \caption{}
\end{subfigure}
\begin{subfigure}[t]{0.44\textwidth}
  \centering
  \includegraphics[width=\textwidth]{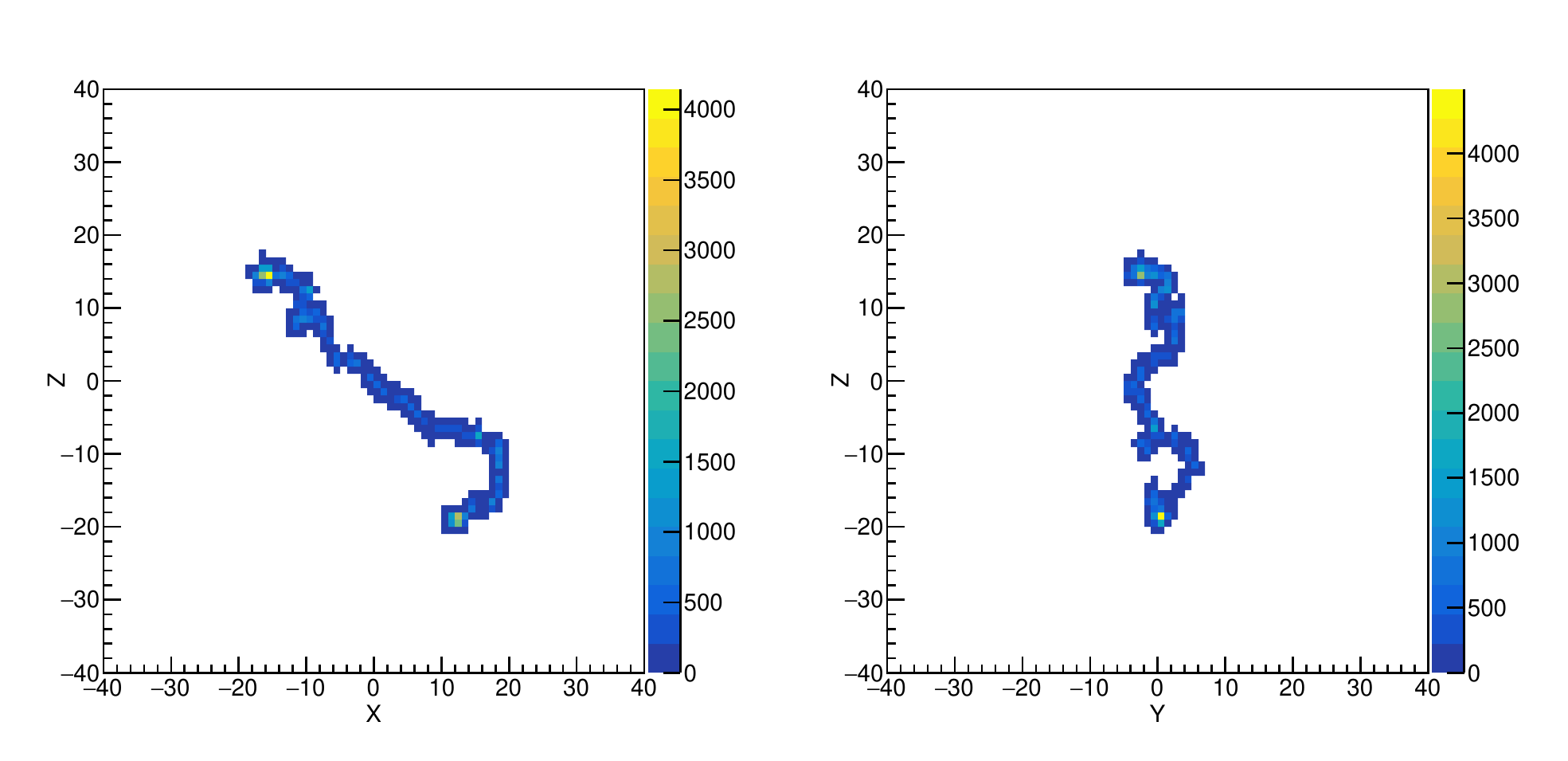}
  \caption{}
\end{subfigure}
\caption{Falsely identified events by CNN in epoch 16. (a) The $x$-$z$
  and $y$-$z$ projection of a NLDBD signal event, which is identified
  as a background event. (b): The $x$-$z$ and $y$-$z$ projection of a
  background event, which is identified as a signal event.}
  \label{fig:missing_event}
\end{figure}

Examples of falsely identified events are given in
Fig.~\ref{fig:missing_event}. The two expected Bragg peaks are not
evident in the miss identified signal events. But in the falsely
identified background, one could easily recognize relative large
energy depositions at the two ends of the tracks.

\begin{table}[hbt]
  \centering
  \begin{tabular}{ccccc}
    \\\hline
    & PandaX-III baseline & CNN (model-16) & CNN (model-18) & CNN (average) \\\hline
    $\epsilon_{s}$ & 0.645 & 0.475  & 0.487 & \\
    $ 1-\epsilon_{b}$ & 0.9714 & 0.9943  & 0.9936 &\\
    $\epsilon_{s}/\sqrt{\epsilon_{b}}$ & 3.816 & 6.264 & 6.098 & 6.174 \\\hline
    improvement & - & $64.2\%$ & $59.8\%$ & $61.8\%$\\\hline
  \end{tabular}
  \caption{Comparison between the results from PandaX-III baseline requirement and the CNN method.}
  \label{tab:comparison}
\end{table}

We also generated a small dataset by simulating the high energy gamma
resulting from the $^{238}$U and $^{232}$Th contamination in the steel
bolts. Nearly identical background rejection power is achieved by
applying the trained network to the dataset. That demonstrates that
the initial position of background gamma has no visible effect on the
discrimination power of CNN.

According to the PandaX-III CDR, the expected background suppression
factor by the traditional topological method is 35.  The comparison
between it and the CNN method is given in
Table.~\ref{tab:comparison}.
The CNN method improves the efficiency
ratio greatly, thus the detection sensitivity will be improved
accordingly.
The large number of parameters in the CNN method helps to
describe the features of physical events better in comparison with the
traditional methods.

\section{Summary}
\label{sec:summary}

We studied the method for the discrimination of signal and background
with CNN in the PandaX-III experiment based on Monte Carlo
simulation. By training a modified ResNet-50 model with digitized MC
data which contains only the $x$-$z$ and $y$-$z$ snapshot information,
we successfully achieved a relatively high efficiency ratio of
$\epsilon_{s,cnn}/\sqrt{\epsilon_{b,cnn}}$, which improves the
corresponding ratio from PandaX-III CDR with topological method by
about $62\%$. Further studies are required to incorporate with the
detector readout response, such as the signal formation for a better
description of the experimental data.

\section*{Acknowledgments}
 This works is supported by the grant from the
Ministry of Science and Technology of China (No. 2016YFA0400302) and
the grants from National Natural Sciences Foundation of China
(No. 11505122 and No. 11775142). We thank the support from the Key
Laboratory for Particle Physics, Astrophysics and Cosmology, Ministry
of Education. This work is supported in part by the Chinese Academy of
Sciences Center for Excellence in Particle Physics (CCEPP).

\appendix

\section{The ambiguity of position reconstruction with strip readout}
\label{sec:limit_strip}
\begin{figure}[tb]
  \centering
  \includegraphics[width=0.4\linewidth]{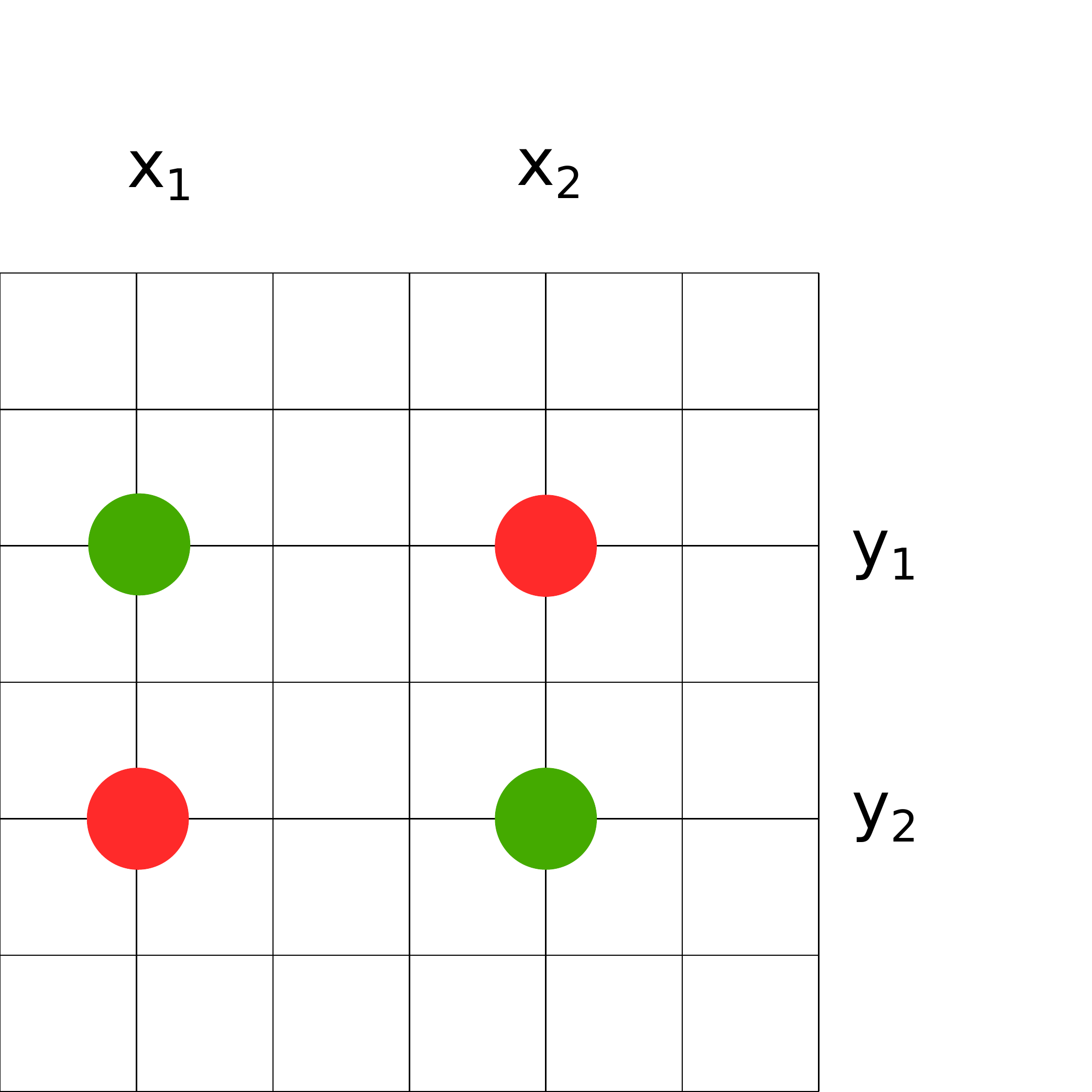}
  \caption{An example of the ambiguity in the position reconstruction with strip readout. The true positions of hits are $(x_1, y_1)$ and $(x_2, y_2)$ (green dots). Because all the $x_1,x_2,y_1,y_2$ strips have been fired, additional fake hits of $(x_1, y_2)$ and $(x_2,y_2)$ may be reconstructed (red dots).}
  \label{fig:strip_limit}
\end{figure}

The MicroMegas modules used by PandaX-III are read out with strips. Simultaneously hits on different strips can be used to reconstructed the positions of the signal. But ambiguity appears when more than 2 strips are fired at the same time, and an example is visualized in Fig.~\ref{fig:strip_limit}.

\section{Comparison between three CNN models}
\label{sec:cnn_comp}
We have tested several different CNN structures for the discrimination of signal and background with a smaller training data set. A comparison between the model complexity, best training accuracy and the signal efficiency at a fixed background rejection efficiency for the testing data is given in Table~\ref{tab:comparison}. The ResNet-50 is finally chosen due to its highest signal efficiency.
\begin{table}[hbt]
  \centering
  \begin{tabular}{cccc}
    \\\hline
    Model & number of trainable parameters & accuracy & $\epsilon_{s}$ \\\hline
    3-Layer Convolutional Model & $720,993$ & $82\%$  & $35.7\%$ \\
    VGG-16\cite{DBLP:journals/corr/SimonyanZ14a} & $15,894,849$ & $92.8\%$  & $73.9\%$ \\
    ResNet-50 & $24,059,393$ & $94.0\%$ & $79.0\%$ \\\hline
  \end{tabular}
  \caption{Simple comparison between different CNN models with a smaller training dataset. The signal efficiency $\epsilon_{s}$ is calculated at a fixed background rejection efficiency of $98.0\%$.}
  \label{tab:comparison}
\end{table}

\end{document}